\documentclass{lmcs} %%% last changed 2014-08-20
\pdfoutput=1

\usepackage{lastpage}
\lmcsdoi{15}{2}{19}
\lmcsheading{}{\pageref{LastPage}}{}{}%
{Oct.~25,~2017}{May~28,~2019}{}


\usepackage{mathtools}
\usepackage{amssymb}

\usepackage{algorithm}
\usepackage{algorithmic}
\newcommand{\ENSUREGAP}{\vspace{0.2cm}}

\DeclareMathOperator{\WL}{WL}
\DeclareMathOperator{\ccu}{cclean}
\DeclareMathOperator{\Aux}{Aux}
\DeclareMathOperator{\val}{val}

\newcommand{\dunion}{\mathbin{\dot\cup}}

\begin{document}

\title[Upper Bounds on FO-Quantifier Depth]{Upper Bounds on the Quantifier Depth for Graph Differentiation in First-Order Logic}
\titlecomment{{\lsuper*}An extended abstract of this paper appeared in the Proceedings of the 2017 32nd Annual ACM/IEEE Symposium on Logic in Computer Science (LICS)}

\author[S.~Kiefer]{Sandra Kiefer}	
\address{RWTH Aachen University, Lehrstuhl Informatik 7, Ahornstra\ss{}e 55, 52074 Aachen, Germany}
\email{kiefer@cs.rwth-aachen.de}  	

\author[P.~Schweitzer]{Pascal Schweitzer}
\address{TU Kaiserslautern, Algorithms and Complexity Group, Postfach 3049, 67663 Kaiserslautern, Germany}	
\email{schweitzer@cs.uni-kl.de}  

%%%%%%%%%%%%%%%%%%%%%%%%%%%%%%%%%%%%%%%%%%%%%%%%%%%%%%%%%%%%%%%%%%%%%%%%%%%

\begin{abstract}
  \noindent We show that on graphs with~$n$ vertices, the~2-dimensional Weisfeiler-Leman algorithm requires at most~$O\big(n^2/\log(n)\big)$ iterations to reach stabilization. This in particular shows that the previously best, trivial upper bound of~$O(n^2)$ is asymptotically not tight. In the logic setting, this translates to the statement that if two graphs of size~$n$ can be distinguished by a formula in first-order logic with counting with~$3$ variables (i.\,e., in~$\mathcal{C}^3$), then they can also be distinguished by a~$\mathcal{C}^3$-formula that has quantifier depth at most~$O\big(n^2/\log(n)\big)$.

To prove the result we define a game between two players that enables us to decouple the causal dependencies between the processes happening simultaneously over several iterations of the algorithm. This allows us to treat large color classes and small color classes separately. As part of our proof we show that for graphs with bounded color class size, the number of iterations until stabilization is at most linear in the number of vertices. This also yields a corresponding statement in first-order logic with counting. 

Similar results can be obtained for the respective logic without counting quantifiers, i.\,e., for the logic~$\mathcal{L}^3$.
\end{abstract}

\maketitle

\section{Introduction}
The Weisfeiler-Leman algorithm is a combinatorial procedure that plays a central role in the theoretical and practical treatment of the graph isomorphism problem. For every~$k$ there is a~$k$-dimensional version of the algorithm, which repeatedly and isomorphism-invariantly refines a partition of the set of~$k$-tuples of vertices of the input graph. This process stabilizes at some point and the final partition can often be used to distinguish non-isomorphic graphs.

On the practical side, the 1-dimensional variant of the algorithm, which is also called color refinement, is an indispensable subroutine in virtually all currently competitive isomorphism solvers (such as~\texttt{Nauty} and~\texttt{Traces}~\cite{mckaypip14}, \texttt{Bliss}~\cite{JunttilaK07} and \texttt{saucy}~\cite{DargaLSM04}). The procedure is also applied to speed up algorithms in other fields, for example in the context of subgraph kernels in machine learning~\cite{ShervashidzeSLMB11} or static program analysis~\cite{LiSSSS16}. Similarly, it can be modified to allow for effective dimension reduction in linear programming~\cite{GroheKMS14}. The stable partition of the~$k$\hbox{-}dimensional Weisfeiler-Leman algorithm can be computed in time~$O(n^{k+1}\log n)$~(see~\cite{BerkholzBG13}), but for practical purposes this running time is often excessive already for~$k=2$.

On the theoretical side, it is known that for random graphs the 1\hbox{-}dimensional Weisfeiler-Leman algorithm asymptotically almost surely correctly decides graph isomorphism~\cite{BabErdSelSta80}. Whereas the~$1$-dimensional algorithm fails to distinguish regular graphs of equal size and equal degree, the~$2$-dimensional version asymptotically almost surely decides isomorphism for random regular graphs~\cite{Kucera87}. While for every graph class with a forbidden minor a sufficiently high-dimensional Weisfeiler-Leman algorithm correctly determines isomorphism~\cite{Grohe12}, it is known that for every~$k$ there are non-isomorphic graphs that are not distinguished by the~$k$-dimensional algorithm~\cite{cfi}. In his recent breakthrough result, Babai~\cite{Babai16} employs a~$\mathrm{polylog}(n)$-dimensional Weisfeiler-Leman algorithm to develop the to date fastest isomorphism algorithm, which solves the graph isomorphism problem in quasi-polynomial time.

We are concerned with the number of iterations required for the Weisfeiler-Leman algorithm to stabilize. 
More specifically, for~$n, k \in \mathbb{N}$ we are interested in~$\WL_k(n)$, the maximum number of iterations required to reach stabilization of the~$k$-dimensional Weisfeiler-Leman algorithm among all (simple) graphs of size~$n$. 
This iteration number plays a crucial role  for the parallelization of the algorithm~\cite{KoblerV08}. 
The trivial upper bound of~$\WL_k(n)\leq n^k-1$ holds for every repeated partitioning of a set of size~$n^k$. For~$k=1$, on random graphs, this iteration number is asymptotically almost surely~$2$~\cite{BabErdSelSta80}, but by considering paths, one quickly determines that~$\WL_1(n)\geq n/2 -1$. This bound was recently improved to~$\WL_1(n)\geq n-O(\sqrt{n})$~\cite{KrebsV15}. For fixed~$k>1$ it is already non-trivial to show linear lower bounds on~$\WL_k(n)$. Modifying the construction of Cai, F\"urer and Immerman~\cite{cfi}, this was achieved by F\"{u}rer \cite{Furer01}~who showed that~$\WL_k(n) \in \Omega(n)$, remaining to date the best known lower bound.

Concerning upper bounds, for~$k \geq 1$, no improvement over the trivial upper bound~$O(n^k)$ has been known so far and for~$k= 1$ it is indeed asymptotically tight. However, for~$k=2$, we show that the trivial upper bound is not tight.

\begin{thm}\label{thm:bd:col:class:main:res}
The number of iterations of the~$2$-dimensional Weisfei\-ler-Leman algorithm on graphs of size~$n$ is at most~$O(n^2/ \log(n))$. 
\end{thm}

There is a close connection between the~$2$-dimensional algorithm and matrix multiplication which is even more prominent in the context of coherent configurations~\cite{BabaiHandbook}. It is possible to execute~$t$ iterations of the~$2$-dimensional Weisfeiler-Leman algorithm by performing~$t$ matrix multiplications over a certain ring (see \cite{chemistry}, Section 5). However, it is also well-known that using randomization, these multiplications can be performed over the integers (see, for example, \cite{schweitzer}, Sections 2.9.2 and 2.9.3), yielding a running time of~$O(n^{\omega})$, where~$\omega < 3$ is the coefficient for
matrix multiplication. See~\cite{Blaser13} for a more general introduction to matrix multiplication.

There is a correspondence between the Weisfeiler-Leman algorithm and the expressive power of first-order logic when it comes to the distinguishability of graphs. We denote by~$\mathcal{L}^k$ the~$k$-variable fragment of first-order logic and by~$\mathcal{C}^k$ the extension of~$\mathcal{L}^k$ by counting quantifiers of the form~$\exists^{\geq i}$. As to the semantics, a graph~$G$ satisfies~$\exists^{\geq i} \, x \varphi$ if there are at least~$i$ vertices~$v \in V(G)$ such that~$G \models \varphi[v/x]$, i.\,e., the graph~$G$ is a model for~$\varphi$ via an interpretation mapping~$x$ to~$v$.\footnote{In the literature, such~$k$-variable logics with or without counting are often defined as fragments of fixed-point logics instead of fragments of first-order logic (see e.g.\ \cite{Otto2017}). For our applications, this distinction is not necessary since we only need formulas for graphs of a fixed size.} For a more detailed introduction to~$\mathcal{L}^k$ and~$\mathcal{C}^k$, we refer the reader to~\cite{immermanlander}. Immerman and Lander showed that two graphs are distinguishable in~$\mathcal{C}^k$ if and only if they can be distinguished by the~$k$\hbox{-}dimensional Weisfeiler-Leman algorithm. In this context, the iteration number of the algorithm corresponds to the quantifier depth of a formula required to distinguish the graphs (see~\cite{MR3418635}), yielding the following corollary.

\begin{cor}\label{cor:logic:dept}
If two~$n$-vertex graphs can be distinguished by the~$3$-variable first-order logic with counting~$\mathcal{C}^3$, there is also a formula in~$\mathcal{C}^3$ with quantifier depth at most~$O(n^2/ \log(n))$ distinguishing the two graphs.
\end{cor}

Finally, there are also certain types of Ehrenfeucht-Fra\"{i}ss\'e games that simulate the Weisfeiler-Leman algorithm and in which the iteration number corresponds to the maximal number of moves in a shortest winning strategy for Spoiler (see~\cite{cfi}). Thus, from upper bounds on the iteration number of the Weisfeiler-Leman algorithm, we also obtain upper bounds on the length of a shortest winning strategy in these games.

\subsection*{Our technique.} Our central technique to prove the upper bound consists in defining a new two-player game mimicking the mechanics of the Weisfeiler-Leman algorithm. While the first player assumes a role of an adversary even stronger than the Weisfeiler-Leman algorithm by being allowed arbitrary refinements, the second player repeatedly rectifies the graph in clean-up steps to maintain various consistency properties for the coloring of the graph. This technique allows us to decouple the causal dependencies between the processes happening simultaneously  over several iterations of the algorithm.
The strategy of the second player takes into account the sizes of the vertex color classes in the current partition. Defining vertex color classes with sizes beyond a certain threshold to be large, we first bound the number of iterations in which color classes are refined that are related in any way with large vertex color classes. We then show that for a fixed threshold, the total amount of iterations dealing exclusively with small vertex color classes is linear. As a consequence we obtain the following lemma.

\begin{lem}\label{lem:linear:for:bounded:col:class}
The number of iterations of the~$2$-dimensional Weisfei\-ler-Leman algorithm on graphs with~$n$ vertices of color class size at most~$t$ is~$O(2^t n)$.
\end{lem}

Similarly to Corollary~\ref{cor:logic:dept}, the lemma also translates into the logic setting yielding linear bounds on the quantifier depth. The graph classes of bounded color class size repeatedly play a role in the context of graph isomorphism. In fact, to show the linear lower bound of~$\WL_k(n)\in \Omega(n)$, F\"{u}rer constructs graphs of bounded color class size~\cite{Furer01}. Even for graphs of bounded color class size, for~$k=2$, prior to this paper, the only available bound was the trivial one of~$\WL_2(n)\leq n^2-1$. For such classes, we now have upper and lower bounds matching up to a constant factor.

Recently, Berkholz and Nordstr\"{o}m were able to obtain a new \emph{lower} bound on the iteration number of the~$k$-dimensional  Weisfeiler-Leman algorithm for finite structures~\cite{BerkholzN16}. Their construction shows that there are pairs of~$n$-element relational structures that are distinguished by the~$k$-dimensional Weisfeiler-Leman algorithm, but not within~$n^{o(k/\log k)}$ refinement steps. This lower bound holds whenever~$k< n^{0.01}$. Since it is close to the trivial upper bound, one might believe that the upper bound is tight.\footnote{In fact, Berkholz's individual result in~\cite{Berkholz14} yields the tightness of the trivial upper bound of~$O(n^4)$ for a logic fragment which is related to our fragments of counting logics, namely for the~$3$-variable existential negation-free fragment of first-order logic.}

However, our results show that for~$k = 2$ this is not the case. Moreover, the construction of Berkholz and Nordstr\"{o}m describes finite structures with bounded color class size. Consequently, it seems that Lemma~\ref{lem:linear:for:bounded:col:class} may be an obstruction for a modification of their structures to graphs with strong lower bounds. Indeed, our theorem says that for graphs of bounded color class size there is a linear upper bound.

\section{Preliminaries}

In this paper, a \emph{colored graph} is a tuple~$G = (V, E, \chi)$ with a finite vertex set~$V$ and edge set~$E = V^2$, in which all edges are assigned colors by the map~$\chi \colon E \rightarrow \mathcal{C}$, i.\,e., they are assigned values from a particular set~$\mathcal{C}$. Note that this definition captures the classical notions of directed colored and uncolored graphs since we can reserve a specific color for non-edges. Furthermore, we can interpret undirected graphs as colored graphs where for all~$u,v \in V$, we have~$\chi(u,v) = \chi(v,u)$. Thus, simple graphs (i.\,e., undirected graphs without loops at vertices) are also special cases of colored graphs with respect to our definition. 

If it is clear from the context, we sometimes simply talk about \emph{graphs} and drop the attribute ``colored''. For the coloring function~$\chi$, we suppose that the set of colors of loops and the set of colors of other arcs are disjoint, that is, we have
\[\big\{\chi(u,u) \mid u \in V \big\} \cap \big\{\chi(u,v) \mid u, v \in V, u \neq v\big\}  = \emptyset.\]

 Let~$G = (V, E, \chi)$ be a colored graph with coloring~$\chi \colon E \rightarrow \mathcal{C}$.

Throughout the paper, we assume that~$\chi(v_1,u_1) = \chi(v_2,u_2)$ if and only if~$\chi(u_1,v_1) = \chi(u_2,v_2)$ for any coloring~$\chi$ we consider. We say the coloring respects \emph{converse equivalence}. (We do not lose generality with this assumption since the 2-dimensional Weisfeiler-Leman algorithm defined below maintains this property of colorings. See Appendix~\ref{app:converse:equiv} for a more detailed discussion.)

The \emph{in-neighborhood} of a vertex~$v \in V$ with respect to a set of colors~$\mathcal{C}' \subseteq \mathcal{C}$ is the set
\[
N^{-}_{\mathcal{C}'}(v) \coloneqq \{u \mid u\in V, \chi(u,v) \in \mathcal{C}'\}.
\]
Likewise, the \emph{out-neighborhood} of~$v$ with respect to~$\mathcal{C}'$ is the set
\[
N^{+}_{\mathcal{C}'}(v) \coloneqq \{u \mid u\in V, \chi(v,u) \in \mathcal{C}'\}.
\]
The \emph{color in-degree} of~$v$ with repect to the set~$\mathcal{C}'$ is the number~$d^{-}_{\mathcal{C}'}(v) \coloneqq |N^{-}_{\mathcal{C}'}(v)|$. The \emph{color out-degree} is defined analogously. When talking about \emph{color degrees}, we mean color in-degrees and color out-degrees.

The coloring~$\chi$ induces a partition~$\pi(\chi)$ of~$V^2$: 
A \emph{color class} of~$G$ is a set of~$2$-tuples that all have the same color. Since we are only interested in the color classes induced by the colorings we consider, and not in the actual colors, we do not distinguish in our notation between a color~$C$ and the corresponding class of~$2$-tuples. 

A \emph{vertex color class} is a color class only consisting of tuples of the form~$(v,v)$ with~$v \in V$. Similarly, an \emph{edge color class} is a color class that consists only of tuples of the form~$(v,w)$ with~$v,w \in V$ and~$v \neq w$. As the term vertex color class already implies, we implicitly identify every vertex tuple~$(u,u)$ with the corresponding vertex~$u$. Consequently, we use the abbreviation~$\chi(u)$ for~$\chi(u,u)$.

For a set~$S \subseteq V^2$ and a vertex color class~$C$, we say that~$S$ is \emph{incident with~$C$} if there exists a~$(u,v) \in S$ with~$(u,u) \in C$ or~$(v,v) \in C$. 

For two partitions~$\pi$ and~$\pi'$, we say that~$\pi'$ is \emph{finer} than~$\pi$ if every element of~$\pi'$ is contained in an element of~$\pi$. We write~$\pi \succeq \pi'$ (and equivalently~$\pi' \preceq \pi$) to express that~$\pi'$ is finer than~$\pi$. Accordingly, we say that~$\pi$ is \emph{coarser} than~$\pi'$.
For a coloring~$\chi$ we denote by~$\pi(\chi)$ the induced partition of~$V^2$. 

For two colored graphs~$G=(V,E,\chi)$ and~$G'=(V',E',\chi')$, we say that~$G'$ \emph{refines}~$G$ (and equivalently, that~$G'$ is a \emph{refinement} of~$G$) if~$V = V'$ and~$\pi(\chi) \succeq \pi(\chi')$. Slightly abusing notation, we write~$G \succeq G'$ in this case. If both~$G \succeq G'$ and~$G' \succeq G$ hold, we write~$G \equiv G'$ and call the two colored graphs~\emph{equivalent}.
Similarly, in the games we define, we say that players \emph{refine} their input graph~$G=(V,E,\chi)$ if they recolor the edges of~$G$ in a such way that the new induced partition of~$V^2$ is finer than~$\pi(\chi)$.

\subsection{The Weisfeiler-Leman algorithm}

By iteratively refining a partition of the set of vertex tuples of its input graph, the Weisfeiler-Leman algorithm computes a stable partition of the graph. For every~$k \in \mathbb{N}$, a~$k$-dimensional variant of the algorithm is defined. We are mainly concerned with the 2-dimensional variant that we describe next. 

\begin{defi}[the~$2$-dimensional Weisfeiler-Leman refinement]\label{def:k:dimensional:weisfeiler:lehman:refinement} Let~$\chi \colon V^2 \rightarrow \mathcal{C}$ be a coloring of the~$2$-tuples of vertices of a graph~$G$, where~$\mathcal{C}$ is some set of colors. We define the \emph{$2$\hbox{-}dimensional Weisfeiler-Leman refinement}~$G^{\textsf{r}}$ to be the graph~$G$ with the coloring~$\chi^{\textsf{r}}$ satisfying~$\chi^{\textsf{r}}(v_1,v_2) = \big(\chi(v_1,v_2); {\mathcal{M}}\big)$ where~${\mathcal{M}}$ is the multiset defined as
\[{\mathcal{M}} \coloneqq \big\{\!\!\big\{\big(\chi(w,v_2), \chi(v_1,w)
\big) \ \big\vert \ w\in V     \big\}\!\!\big\}.\]
\end{defi}

It is not difficult to see that~$\pi(\chi) \succeq \pi(\chi^{\textsf{r}})$. 

For a colored input graph~$G$, we set~$G^{(0)} \coloneqq G$ and~$G^{(i)} \coloneqq (G^{(i-1)})^{\textsf{r}}$ for~$i \geq 1$. Executing~$i$ \emph{iterations of the 2-dimensional Weisfeiler-Leman algorithm} means computing~$G^{(i)}$. The \emph{output} of the~$2$-dimensional Weisfeiler-Leman algorithm on input~$G$ is~$G^{(k)}$, where~$k$ is the smallest integer satisfying~$G^{(k)} \equiv G^{(k+1)}$. We denote this graph by~$\widetilde{G}$. We call~$G$ and the induced partition of the~$2$-tuples of its vertices \emph{stable} if~$G = \widetilde{G}$. Accordingly, we call~$\widetilde{G}$ the \emph{stabilization} of~$G$. 

The stable partitions of the set of arcs under the 2-dimensional Weisfeiler-Leman algorithm are in fact the coherent configurations over the vertex set that respect the edge relation. For an extended background on the theory of coherent configurations, we refer for example to~\cite{MR1994960}.

The~$2$-dimensional Weisfeiler-Leman algorithm can be used to check whether two given (colored or uncolored) graphs are non-isomorphic by computing the stabilizations and rejecting if there is a color~$C$ such that the numbers of~$C$-colored vertex pairs in the two graphs differ. However, even if the stabilizations of the two graphs agree in every number of vertices for a particular color, the graphs might not be isomorphic. It is not trivial to describe for which graphs this isomorphism test is always successful (see~\cite{KieferSS15}).

In the following, we drop the dimension, i.\,e., when we talk about \emph{the Weisfeiler-Leman algorithm} or \emph{refinement}, we always mean the~$2$-dimensional variant.

\section{An upper bound on the iteration number}

Now we prove the upper bound on the number of iterations of the Weisfeiler-Leman algorithm. That is, we prove Theorem~\ref{thm:bd:col:class:main:res} showing that on any colored graph with~$n$ vertices, the~$2$-dimensional Weisfeiler-Leman algorithm terminates after~$O(n^2/\log n)$ iterations. 

For this, we are going to define a game in which two players alternate in their turns on a colored graph. We show that the costs of the game form an upper bound on the iteration number of the Weisfeiler-Leman algorithm. Therefore, to prove Theorem~\ref{thm:bd:col:class:main:res} it suffices to show that the costs of the game, assuming the players play optimally, are~$O(n^2/\log n)$.

The game represents a sequential way of looking at the Weisfeiler-Leman algorithm. Since one turn in the game consists in elementary actions, which can be performed one after the other and not all at the same time as in one iteration of the Weisfeiler-Leman algorithm, the costs of the game are easier to analyze.

\subsection{Description of the game}
The game starts with a colored graph on vertex set~$V$. Player 1 begins and the players alternate turns. Each turn consists in choosing a refinement of the current coloring of the graph. If in his turn Player 1 is faced with the colored graph~$G$, then he must choose a proper refinement of~$G$, that is, he must return a graph~$G'$ with~$G \succneqq G'$. In contrast to this, in a turn of Player 2, when faced with the graph~$G$, she has to return a graph~$G'$ with~$G\succeq G' \succeq \widetilde{G}$. Thus, the refinement that Player~$2$ chooses must still be coarser than~$\widetilde{G}$, but it may be equivalent to~$G$ or to~$\widetilde{G}$.
The game ends when the coloring of the graph induces the discrete partition, i.\,e., the partition~$\{\{(u,v)\} \mid u,v \in V\}$. 

Each turn of Player~$1$ adds 1 to the total costs of the game, whereas the costs for a turn of Player~$2$ playing~$G'$ as a response to~$G$ is the smallest integer~$j$ such that~$G'\succeq G^{(j)}$. Thus, a turn of Player~$2$ costs the number of iterations of the Weisfeiler-Leman algorithm that she needs to perform in order to obtain a refinement of the graph she returns to Player~$1$. In the game, every modification of a graph that increases the total costs by~$1$ is called a \emph{move}. Note that the game has costs of at most~$O(n^2)$.

Player~$1$ aims at maximizing the total costs, whereas his opponent Player~$2$ wants to minimize it. Given a strategy $S_1$ of Player~$1$ and a strategy $S_2$ of Player~$2$, let $\val(S_1,S_2)$ denote the costs of the game when Player~$1$ uses $S_1$ and Player~$2$ uses $S_2$. Since the game is finite and deterministic and has perfect information, it is not hard to see that the value $c(G)$ of the game (i.\,e., the costs assuming both players play optimally) is well-defined and that
\[\max_{S_1} \min_{S_2} \val(S_1, S_2) = c(G) = \min_{S_2} \max_{S_1} \val(S_1, S_2).\]

Note that we can also view the game as a zero-sum game (when considering the costs as the gain for Player 1 and the loss for Player 2). 

The following lemma states an important monotonicity fact for our game.

\begin{lem}\label{lem:monoton}
 Let~$G$ and~$H$ be two colored graphs with~$G \succeq H$. Then the following hold.
 \begin{enumerate}
  \item~$G^{(i)} \succeq H^{(i)}$ for every~$i$.\label{item:one}
  \item~$\widetilde{G} \succeq \widetilde{H}$. \label{item:two}
  \item If additionally~$H \succeq \widetilde{G}$, then for every~$i$ with~$G^{(i)} = \widetilde{G}$, we have that~$H^{(i)} \equiv \widetilde{G}$. In particular~$\widetilde{H} \equiv \widetilde{G}$. \label{item:three}
 \end{enumerate}
\end{lem}

\begin{proof}
 Part~\ref{item:one} follows straight from Definition~\ref{def:k:dimensional:weisfeiler:lehman:refinement} using induction on~$i$: For~$i=0$, the statement is the assumption. Suppose that~$G^{(i)} \succeq H^{(i)}$ for a fixed~$i$. Let~$\chi_G$ and~$\chi_H$ denote the colorings in~$G^{(i)}$ and~$H^{(i)}$, respectively, and let~$\chi_G^{\textsf{r}}$ and~$\chi_H^{\textsf{r}}$ denote the colorings in~$G^{(i+1)}$ and~$H^{(i+1)}$. Let~$V \coloneqq V(G) = V(H)$. Let~$(u_1,u_2)$ and~$(v_1,v_2)$ be two tuples with~$u_1, u_2, v_1, v_2 \in V$ and such that~$\chi_G^{\textsf{r}}(u_1,u_2) \neq \chi_G^{\textsf{r}}(v_1,v_2)$. We show that~$\chi_H^{\textsf{r}}(u_1,u_2) \neq \chi_H^{\textsf{r}}(v_1,v_2)$. 

 If~$\chi_G(u_1,u_2) \neq \chi_G(v_1,v_2)$, the statement follows from~$G^{(i)} \succeq H^{(i)}$ and~$\pi(\chi_H) \succeq \pi(\chi_H^{\textsf{r}})$. Otherwise~$\chi_G(u_1,u_2) = \chi_G(v_1,v_2)$ and therefore, it must hold that 
 \[\big\{\!\!\big\{\big(\chi_G(w,u_2), \chi_G(u_1,w)
\big) \ \big\vert \ w\in V     \big\}\!\!\big\} \neq 
\big\{\!\!\big\{\big(\chi_G(w,v_2), \chi_G(v_1,w)
\big) \ \big\vert \ w\in V     \big\}\!\!\big\}.\]

 However, since~$G^{(i)} \succeq H^{(i)}$, the same inequality holds when substituting~$\chi_H$ for~$\chi_G$. Thus, the tuples~$\chi_H^{\textsf{r}}(u_1,u_2)$ and~$\chi_H^{\textsf{r}}(v_1,v_2)$ differ in their second components and hence, they are distinct.

 Thus, if~$i$ is the minimal integer with~$G^{(i)} = \widetilde{G}$, then it holds that~$\widetilde{G} \succeq H^{(i)} \succeq \widetilde{H}$, which proves Part~\ref{item:two}. Part~\ref{item:three} is an immediate consequence of Part~\ref{item:one}, using that~$G^{(i)} \succeq H^{(i)} \succeq \widetilde{G}$.
\end{proof}

Hence, the game is monotone in the following sense: Given an input graph~$H$, the costs for Player~$2$ to obtain the stabilization~$\widetilde{H}$ are at most as high as the costs for her to obtain the stabilization of any other input graph~$G$ with~$G \succeq H$ and~$\widetilde{G} \equiv \widetilde{H}$.

We first show that in an optimal strategy, Player~$2$ does not
execute partial iterations of the Weisfeiler-Leman algorithm on her input. 

\begin{lem}\label{lemma_pl2}
 Suppose that in one of her turns, Player~$2$ is given a graph~$G$. If she plays optimally, she returns~$G^{(i)}$ for some integer~$i$.
\end{lem}

\begin{proof}
Consider a strategy~$S$ for Player~$2$ which dictates to her for at least one input~$G$ for her turn to play a graph~$G'$ such that~$G' \not\equiv G^{(j)}$ for every~$j$. In the game tree~$T$ with this fixed strategy for Player~$2$, consider a counterexample situation that is maximal with respect to the number of moves performed so far. That is, consider a node~$x \in V(T)$ at which, on input~$G$, Player~$2$ finishes her turn with a graph~$G'$ such that~$G' \not\equiv G^{(j)}$ for every~$j$ and for which the subtree of~$T$ rooted at~$x$ contains no such node. Thus, in her future turns, Player~$2$ always returns a graph corresponding to a full iteration of the Weisfeiler-Leman algorithm. 

Let~$i$ be the maximal~$j$ with~$G^{(j)} \succneqq G'$. Then Player~$1$ can define~$G^{(i+1)}$ as the new input for Player~$2$. By the maximality of the counterexample, Player~$2$ now finishes this turn with a graph~$G^{(i+\ell)}$ with~$\ell \geq 1$. This conversion of~$G$ into~$G^{(i+\ell)}$ contributes the costs~$(i+1) + 1 + \big(i+\ell-(i+1)\big) = i+\ell+1$, whereas Player~$2$ could have reached~$G^{(i+\ell)}$ with~$i+\ell$ moves by simply executing the~$i+\ell$ necessary iterations. Thus, the strategy~$S$ is not optimal. Hence, in every optimal strategy Player~$2$ only returns graphs corresponding to full iterations of the Weisfeiler-Leman algorithm.
\end{proof}

It is even optimal for Player~$2$ to return the stabilization of her input.

\begin{cor}\label{cor_p2}
Given a colored graph~$G$, it is optimal for Player~$2$ to return~$\widetilde{G}$. 
\end{cor}

\begin{proof}
 Similarly to the proof of Lemma~\ref{lemma_pl2}, consider the counterexamples in which Player~$2$ is given a graph~$G$ and finishes her turn with a graph~$G^{(i)} \succneqq \widetilde{G}$ and it would be worse for her to return~$\widetilde{G}$ instead. Take such a counterexample situation that is maximal with respect to the number of moves performed so far. Given~$G^{(i)}$, Player~$1$ can define~$G^{(i+1)}$ as the new input for Player~$2$ (including the case that~$i+1 = k$). The conversion costs~$i$ for the moves of Player~$2$ and~$1$ for Player~$1$, so altogether~$i+1$. By the maximality of the counterexample, from now on, we can assume that Player~$2$ always returns the stabilization of her input. If~$i+1 = k$, that is~$G^{(i+1)} = \widetilde{G}$, the costs are the same as if Player~$2$ had chosen to stabilize~$G$ herself. If~$i+1 < k$, Player~$2$ now needs~$k-(i+1)$ additional moves to stabilize~$G^{(i+1)}$, because~$\widetilde{G^{(i+1)}} = \widetilde{G}$ by Lemma~\ref{lem:monoton}.
 So the conversion costs are~$i+1+\big(k-(i+1)\big) = k$.
 That is, in all cases, Player~$1$ can choose to pursue a strategy such that the costs are at least as high as when Player~$2$ always returns stable graphs.
\end{proof}

To see that the total costs of the game yield an upper bound on the number~$\WL(G) \coloneqq \WL_2(G)$ of iterations of the Weisfeiler-Leman algorithm on input~$G$, it suffices to describe a strategy for Player~$1$ with which the game has costs at least~$\WL(G)$. For this, Player~$1$ can simply recolor the arcs according to the first iteration of the Weisfeiler-Leman algorithm and define the graph~$G^{(1)}$ as the first input for Player~$2$. Corollary~\ref{cor_p2} states that it is optimal for Player~$2$ to perform the entire Weisfeiler-Leman algorithm on her input. Thus, she could not do better but to perform the remaining iterations of the algorithm on~$G^{(1)}$, resulting in costs that are at least~$\WL(G)$. (They could be higher because the game does not end before the discrete partition is reached.)

We therefore obtain the following corollary, which allows us to analyze the costs of the described game in order to obtain an upper bound on the number of iterations of the~$2$-dimensional Weisfeiler-Leman algorithm.

\begin{cor}
Let~$G$ be a graph and let~$c(G)$ be the value of the~$2$-player game on input~$G$. Then~$\WL_2(G) \leq c(G)$.
\end{cor} 

As a consequence of Corollary~\ref{cor_p2}, we can draw conclusions about optimal strategies for Player~$1$.

\begin{lem}
If playing~$G'$ in response to~$G$ is optimal for Player~$1$ then playing~$G''$ with~$G \succneqq G'' \succeq G'$ is also optimal for Player~$1$.
\end{lem}

\begin{proof}
 Given a colored input graph~$G$, a turn of Player~$1$ consists in choosing a set~$\mathcal{C}'$ of color classes in~$G$ that he wants to refine. We denote by~$G_{\mathcal{C}'}$ the graph that Player~$1$ returns to Player~$2$. 
 
 We show that it is optimal for Player~$1$ to refine only one color class per turn. Suppose Player~$1$ is given a graph~$G$ and decides to refine a set~$\mathcal{C}'$ consisting of~$\ell$ color classes, i.\,e.,~$\mathcal{C}' = \{ C_i \mid 1 \leq i \leq \ell\}$ with~$\ell > 1$. 

 Let~$k$ be the minimal number with~$\widetilde{G}_{\mathcal{C}'} = G^{(k)}_{\mathcal{C}'}$. Due to Corollary~\ref{cor_p2}, we can assume Player~$2$ always returns the stabilization of her input. Hence, the costs for Player~$2$ to return~$\widetilde{G}_{\mathcal{C}'}$ are~$k$.
 We show by induction that the costs do not decrease when Player~$1$ only refines one color class per turn. 
 
 Suppose Player~$1$ plays the graph~$G_{C_1}$, which only refines~$C_1$. Player~$2$ will then stabilize~$G_{C_1}$ with, say,~$j$ iterations. That is, she returns~$\widetilde{G}_{C_1} = G^{(j)}_{C_1}$. From Part~\ref{item:one} of Lemma~\ref{lem:monoton}, we obtain that~$\widetilde{G}_{C_1} \succeq G^{(j)}_{\mathcal{C}'}$. If~$j \geq k$, Player~$1$ can refine~$\widetilde{G}_{C_1}$ to the graph~$\widetilde{G}_{\mathcal{C}'}$ and continue as if~$\widetilde{G}_{\mathcal{C}'}$ was his input graph for this round. 
 
Assume therefore that~$j < k$.
We know that\[\widetilde{G}_{C_1} = G^{(j)}_{C_1} \succeq G^{(j)}_{\mathcal{C}'} \succeq G^{(k)}_{\mathcal{C}'} = \widetilde{G}_{\mathcal{C}'}.\] The second ``$\succeq$'' is proper since~$j < k$ and~$k$ is minimal. From Part~\ref{item:three} of Lemma~\ref{lem:monoton}, we know that the stabilization of~$G^{(j)}_{\mathcal{C}'}$ is~$\widetilde{G}_{\mathcal{C}'}$. This implies that the first ``$\succeq$'' is also proper because~$\widetilde{G}_{C_1}$ is stable whereas~$G^{(j)}_{\mathcal{C}'}$ is not.

This shows that~$\widetilde{G}_{C_1}$ is strictly coarser than~$G^{(j)}_{\mathcal{C}'}$. Thus, Player~$1$ can proceed by refining~$\widetilde{G}_{C_1}$ to~$G^{(j)}_{\mathcal{C}'}$, on which Player~$2$ needs~$k-j$ moves to obtain the stabilization.

In a similar way, one can show that it is optimal for Player~$1$ to split the selected color class only into two color classes.
\end{proof}

From now on, we can thus assume that Player 1 only refines one color class per turn and that he splits this class into exactly two new color classes. What is the effect of such a turn of Player~$1$? By recoloring loops and edges, Player~$1$ refines vertex color classes and edge color classes. As a consequence, some regularity conditions may no longer hold in the obtained graph~$G$. We are in particular interested in the following two conditions.

\begin{enumerate}[label=(C\arabic*),ref=C\arabic*,align=left,labelwidth=\widthof{(C2)},leftmargin=\widthof{(C2) \ }]
 \item  The color of an edge determines the color of its head and of its tail. More precisely, for~$u,v \in V(G)$ the color of~$(u,v)$ determines both the colors of~$(u,u)$ and~$(v,v)$.\label{item:c1} 
 
 \item For every edge color class~$C$ and every two vertex color classes~$C_1$ and~$C_2$, the graph which is induced by~$C$ between~$C_1$ and~$C_2$ is biregular. By this we mean that for every two vertices~$v_1, v_2 \in C_1$ we have~$|N^+_C(v_1) \cap C_2| = |N^+_C(v_2) \cap C_2|$ and~$| N^-_C(v_1) \cap C_2 | = | N^-_C(v_1) \cap C_2|$ and similarly for the~$C$-neighborhoods in~$C_1$ of vertices in~$C_2$.\label{item:c2}
\end{enumerate}

In her reaction to a turn of Player~$1$, we let Player~$2$ ``clean up'' the graph: By refining edge color classes, first she reestablishes Condition~\ref{item:c1} and after that Condition~\ref{item:c2}, i.\,e., the color-biregularity. 
We claim that the costs of this procedure only amount to~$2$. Indeed, to satisfy Condition~\ref{item:c1}, Player~$2$ can simply perform one iteration of the Weisfeiler-Leman algorithm on her input and ignore various refinements that would be made in this iteration. More precisely, given~$G$ colored with~$\chi$ she can return the coloring that assigns~$(u,v)$ the color\[\big(\chi(u,v),\chi(v,v),\chi(u,u)\big).\]

Thus, in the first move of her turn, she only refines (real) edge color classes whose heads (and tails, respectively) are not all of the same color. In the second move, she reestablishes color regularity, i.\,e., she refines vertex color classes according to their color degrees. We call these two moves a \emph{clean-up step}.

It is possible that the attempt to establish Condition~\ref{item:c2} causes Condition~\ref{item:c1} to be violated again. Indeed, if the move splits vertex color classes, Condition~\ref{item:c1} might need to be reestablished afterwards due to the splitting, i.\,e., Player~$2$ might need to continue cleaning-up. In turn this might cause Condition~\ref{item:c2} to be violated, and so on.
We call a shortest succession of clean-up steps that reestablishes both Condition~\ref{item:c1} and Condition~\ref{item:c2} simultaneously a \emph{complete clean-up step}. We denote by~$\ccu(G)$ the colored graph resulting from a complete clean-up step on~$G$.

The following observation allows Player~$2$ to always perform complete clean-up steps without causing critical extra costs.

\begin{obs}\label{obs:vertexsplits}
There are only~$O(n)$ splittings of vertex color classes in the game.
\end{obs}

A clean-up step that is not yet a complete clean-up step consists of at least one splitting of a vertex color class. Therefore, the total costs for a complete clean-up step can be bounded by~$2s + 2 = O(s)$, where~$s$ is the number of vertex color class splittings appearing in the complete clean-up step. (The additional~$2$ is for the costs of the last clean-up step, which does not need to incur a vertex color class splitting.) By Observation~\ref{obs:vertexsplits}, there are only~$O(n)$ vertex color class splittings in the whole game and each of them appears in at most one complete clean-up step. We obtain the following bound on the costs for clean-up steps that split vertex color classes.

\begin{cor}
The total costs for complete clean-up steps that split vertex color classes amount to~$O(n)$.
\end{cor}

A complete clean-up step in which no vertex color classes are split consists only of one clean-up step. Therefore, such a complete clean-up step has constant costs of at most~$2$. This means that we can assign these costs to the preceding move (either performed by Player~$1$ or Player~$2$) since every move entails at most one complete clean-up step. Thus, the extra costs for complete clean-up steps do not have an effect on the asymptotic bound that we want to obtain for the costs of the game.

Now we know optimal strategies for both players and we have seen that Player~$2$ can perform complete clean-up steps whenever she wants without asymptotically increasing the costs. We can thus assume that Player~$2$ always performs a complete clean-up step on her input graph before and after manipulating it.

\subsection{Large color classes}

To give a bound on the costs of the game, we distinguish between large vertex color classes and small vertex color classes with respect to some threshold function~$t$. 

Let~$t \colon \mathbb{N} \rightarrow \mathbb{N}$ be a function with~$t(n) > 0$ for every~$n$.
We call a vertex color class of~$G$ \emph{large} with respect to~$t$ if it consists of at least~$t(n) = t(|V|)$ vertices. A \emph{small} vertex color class is one that is not large.

The choice of the threshold function~$t$ for our purposes will be a tradeoff between the costs for operations on edges incident with large vertex color classes and edges incident with small vertex color classes. We will later see that setting~$t(n) \coloneqq \log_2(n)/2$ suffices to prove Theorem~\ref{thm:bd:col:class:main:res}.

We first treat edges that are incident with large vertex color classes. Due to Observation~\ref{obs:vertexsplits}, we only have to consider moves in which no vertex color classes are split.

For a coloring~$\chi$, we define a potential function by setting
\[f(\chi) \coloneqq \sum\limits_{v \in V} { \lvert \{ \chi(v,w) \mid w \in V \} \rvert }.\]
Obviously,~$f(\chi) \leq n^2$ for all colorings~$\chi$ and~$f$ is strictly monotonically increasing. That is, if~$\pi(\chi) \succneqq \pi(\chi')$ then~$f(\chi) <f(\chi')$.

Let~$B_1, B_2, \dots$ be an enumeration of the large vertex color classes and let~$B \coloneqq \dot{\bigcup}_k {B_k}$ be the set of vertices in large vertex color classes. 

\begin{lem}
Suppose that the current coloring in the game is~$\chi$ and that the current player chooses a refinement which for some~$v \in B$ induces a strictly finer partition on~$\{ (v,w) \mid w \in V\}$ than~$\chi$. If a subsequent complete clean-up step does not split any large vertex color class, then~$f$ increases by at least~$t(n)$.
\end{lem}

\begin{proof}
Let~$v$ be as in the assumptions of the lemma and let~$B_k$ be the large vertex color class that contains~$v$. Denote by~$\chi'$ the coloring obtained after a subsequent clean-up step.  
Then in particular there is a certain, not necessarily large, vertex color class~$C$ such that on the set~$\{ (v,w) \mid w \in C\}$, the coloring~$\chi'$ induces a strictly finer partition than~$\chi$.
Thus, the number of color classes occurring in the set~$\{ (v,w) \mid w \in C\}$ increases by at least~$1$. 

By assumption, the vertex color class~$B_k$ is not split in this round and we can assume Player~$2$ always performs complete clean-up steps. Therefore, to ensure all vertices in~$B_k$ have identical color degrees (Condition~$2$), also on every set~$\{ (v',w) \mid w \in C\}$ with~$v' \in B_k$, the coloring~$\chi'$ must induce a partition strictly finer than the one induced by~$\chi$. Hence, for each of these sets, of which there are at least~$t(n)$, the number of occurring color classes  increases by at least~$1$.

For any vertex~$v \in V$ and any vertex set~$V'$, we have~$\lvert \{ \chi' (v,w) \mid w \in V' \} \rvert \geq \lvert \{ \chi (v,w) \mid w \in V' \} \rvert$ and so
\begin{align*}
f(\chi') &= \sum\limits_{v \in V} { \lvert \{ \chi' (v,w) \mid w \in V \} \rvert } 
\\&= \sum\limits_{v \in B_k} { \lvert \{ \chi' (v,w) \mid w \in V \} \rvert }  +
 \sum\limits_{v \in V \backslash B_k} { \lvert \{ \chi' (v,w) \mid w \in V \} \rvert }
\\&\stackrel{\text{C1}}{=} \sum\limits_{v \in B_k} { \lvert \{ \chi' (v,w) \mid w \in C \} \rvert }
+ \sum\limits_{v \in B_k} { \lvert \{ \chi' (v,w) \mid w \in V \backslash C \} \rvert } + {}
\\&\phantom{=}\sum\limits_{v \in V \backslash B_k} { \lvert \{ \chi' (v,w) \mid w \in V \} \rvert }
\\&\geq  \sum\limits_{v \in B_k} { \lvert \{ \chi (v,w) \mid w \in C \} \rvert } + t(n)
 + \sum\limits_{v \in B_k} { \lvert \{ \chi (v,w) \mid w \in V \backslash C \} \rvert } + {}
\\&\phantom{=}\sum\limits_{v \in V \backslash B_k} { \lvert \{ \chi (v,w) \mid w \in V \} \rvert }
\\&\geq t(n) + f(\chi)
\end{align*}

and this gives the claim.
\end{proof}

\begin{cor}\label{cor:large:color:classes}
The costs for the moves (and the following complete clean-up steps) in which the color set of edges incident with large vertex color classes is properly refined are~$O\big(n^2/t(n)\big)$.
\end{cor}

This bound already includes the necessary complete clean-up steps. In the following, we treat the moves in which edge color classes that are only incident with small vertex color classes are refined. 
 
\subsection{Small color classes} 
 
To analyze the costs for small vertex color classes, we describe a strategy for Player~$2$ on them. For this, we define auxiliary graphs which she uses to derive moves in the original game. For a colored graph~$G$ from the original game, its auxiliary graph is denoted by~$\Aux(G)$. Just like with the clean-up steps in the original game, Player~$2$ will pursue a strategy according to information obtained from auxiliary graphs in order to maintain certain invariants. 

\begin{nota}
We describe all graphs coming from the original game using the letter~$G$ and all auxiliary graphs using the letter~$H$ or, when derived from a specific graph~$G$, by~$\Aux(G)$, in order to distinguish between them more clearly.
\end{nota}

Every auxiliary graph is undirected, simple, uncolored and not necessarily complete. For the sake of readability, we consider the edge set of an auxiliary graph~$H$ with vertex set~$V(H)$ to just be a subset of~$\{\{u,v\} \mid u,v \in V(H), u \neq v\}$ and we drop the notion of a coloring for auxiliary graphs. Whereas in the original game, a move \emph{recolors} edges, in the auxiliary graphs edges are \emph{inserted}.

Next we describe how to obtain the auxiliary graph for a graph~$G$ appearing during a run of the game. The auxiliary graph~$\Aux(G)$ is constructed as follows: Let~$G_1,G_2,G_3,\ldots,G_{\ell} =G$ be the graphs that were played by the players so far. Let~$\mathcal{T}$ be the collection of vertex sets~$C\subseteq V(G)$ that each form a small vertex color class in some~$G_i$.

The vertices of the auxiliary graph~$\Aux(G)$ form a partition into two sets called the \emph{upper} and the \emph{lower vertices} ($V_u$ and~$V_\ell$, respectively). They are two identical copies of the set defined as
\[
\big\{ (C,M) \mid C\in \mathcal{T}, M \subseteq C\big\},
\]
that is, the set that contains all pairs of a small vertex color class~$C$ that has appeared so far in the game and a subset of~$C$.
Thus, as the game progresses the auxiliary graphs have more and more vertices. 

For every small vertex color class~$C\in \mathcal{T}$, the graph~$\Aux(G)$ contains~$2^{|C|}$ upper vertices and also~$2^{|C|}$ lower vertices. Thus, there are at most~$4n \cdot 2^{t(n)}$ vertices in~$\Aux(G)$. (To see this, one observes that there are at most~$2n$ elements in~$\mathcal{T}$.) We have an undirected edge between an upper vertex~$(C,M)\in V_u$ and a lower vertex~$(D,N)\in V_\ell$ if there exists a set of colors~$\mathcal{C}' \subseteq \mathcal{C}$ such that in~$G$, every vertex~$v \in C$ satisfies
\[
v \in M \iff N^{+}_{\mathcal{C}'}(v) = N.
\]

This means that all the edges between~$M$ and~$N$ in~$G$ have a color contained in~$\mathcal{C}'$. Another way of formulating this is that the vertices of~$M$ are exactly the ones whose~$\mathcal{C}'$-neighborhood is~$N$.

Between two upper vertices~$(C,M)$ and~$(C',M')$, we insert an undirected edge if such a condition holds in both directions---more precisely, if there are sets of colors~$\mathcal{C}', \mathcal{C}'' \subseteq \mathcal{C}$ such that every vertex~$v \in C'$ satisfies
\[
v \in M \iff N^{+}_{\mathcal{C}'}(v) = M',
\]
and every vertex~$v \in C''$ satisfies
\[
v \in M' \iff N^{+}_{\mathcal{C}''}(v) = M.
\]

The resulting graph is the auxiliary graph~$\Aux(G)$.
\medskip

\begin{rem}\label{rem:aux}
 Let~$G$ and~$G'$ with~$G \succeq G'$ be two graphs appearing during the original game (i.\,e., the graph~$G'$ appears later than~$G$). Then it holds that~$\Aux(G')\supseteq \Aux(G)$, since by definition, the graph~$\Aux(G')$ contains all the vertices and all the edges from previous auxiliary graphs.
\end{rem}

To describe the strategy that Player~$2$ will pursue in the original game, we need the notion of a \emph{triangle completion}, an operation on undirected uncolored graphs, which we define next. Whereas the clean-up steps are designed to simulate information essentially collected by the~$1$-dimensional Weisfeiler-Leman algorithm, the triangle completion is designed to capture certain dynamics of the~$2$-dimensional algorithm, as will become apparent.

For an undirected uncolored graph~$H$ with~$V(H) = V_u \dunion V_\ell$, its \emph{triangle completion}~$\triangle(H)$ is obtained by applying the following rules once.

\begin{itemize}
 \item Insert an edge between every two upper vertices that have a common neighbor in~$V_\ell$ or in~$V_u$.
 \item Insert an edge between every upper vertex and every lower vertex that have a common neighbor in~$V_u$.
\end{itemize}

Note that no edges between vertices in~$V_\ell$ are inserted. On the obtained graph~$\triangle(H)$, new applications of the above rules may be possible. The graph obtained after~$i$ repetitions of the triangle completion on~$H$ is denoted by~$\triangle^i (H)$. We call~$H$ \emph{stable} if~$\triangle(H) = H$. 

\begin{rem}\label{remark:stable}
An undirected uncolored graph~$H$ with~$V(H) = V_u \dunion V_\ell$ is stable if and only if every connected component of~$H$ has the following properties:
\begin{itemize}
 \item The graph induced on~$V_u$ is a clique.
 \item The graph induced by the edges between~$V_u$ and~$V_\ell$ is complete bipartite.
\end{itemize}
\end{rem}

Now we describe the strategy that Player~$2$ derives from the auxiliary graphs for the original game. It is also shown in Algorithm~\ref{algo}. Player~$2$ first performs a complete clean-up step on her input graph. Then as long as the auxiliary graph of her obtained graph is not stable, she iterates the following process on it: First she performs an iteration of the Weisfeiler-Leman algorithm on~$G$, then a complete clean-up step. She returns the first completely cleaned-up graph~$G'$ for which~$\Aux(G')$ is stable to Player~$1$ who then performs his next turn. 

\begin{algorithm}[H]

\caption{One turn of Player~$2$ in the~$2$-player game on input~$G$.}
\label{algo}
\begin{algorithmic}[1]
\REQUIRE A colored graph~$G$.
\ENSURE A refinement~$G'$ of~$G$ satisfying~$G\succeq G'\succeq \widetilde{G}$.
\ENSUREGAP
\STATE~$G \leftarrow \ccu(G)$\label{line:computed:cleaned:graph:first}
\WHILE {$\triangle\big(\Aux(G)\big) \neq \Aux(G)$}
	\STATE~$G \leftarrow G^{(1)}$
	\STATE~$G \leftarrow \ccu(G)$ \label{line:computed:cleaned:graph}
\ENDWHILE
	\RETURN $G$
\end{algorithmic}
\end{algorithm}

We claim that if Player~$2$ follows this strategy, then she only computes~$O(n \cdot 2^{t(n)})$ pairwise different auxiliary graphs. To show this, we need the following lemma.

\begin{lem}\label{lemma:aux}
 Let~$H^1, \dots, H^k$ be a sequence of graphs such that for all~$i$, the following hold.
 \begin{itemize} 
  \item $V(H^i) = V_\ell^i \dunion V_u^i$.
  \item $|V_\ell^i| = |V_u^i| \leq m$ for some~$m$.
  \item $V_\ell^{i+1} \supseteq V_\ell^i\ $ and $\ V_u^{i+1} \supseteq V_u^i$.
  \item $\triangle(H^i) \subseteq H^{i+1}$.
  \item $H^i \neq H^{i+1}$.
 \end{itemize}
Furthermore assume that the graph induced by each~$V_\ell^i$ is empty. Then~$k \in O(m)$.
\end{lem}

\begin{proof}
 Without loss of generality, assume~$|V_\ell^i| = m$ for every~$i$. Thus, we may assume that the vertices of all the~$H^i$ are the same set~$V_u \dunion V_\ell$.
  For a fixed~$k$, let~$H^1, H^2, \dots, H^k$ be a minimal counterexample to the statement of the lemma, i.\,e., choose the graphs~$H^i$ to have as few edges as possible. This implies that for the cases that~$\triangle(H^i) \neq H^i$ we have~$H^{i+1} = \triangle(H^i)$ and for every other~$i$ we have~$|E(H^{i+1}) \backslash E(H^i)| = 1$. We may assume that~$H^1$ is empty. 
 
 We claim that for any~$j$ with~$\triangle (H^j) = H^j$, there is some~$r \leq 4$ such that~$\triangle(H^{j+r}) = H^{j+r}$, i.\,e., the graph~$H^{j+r}$ is stable again.
 
 Suppose~$\triangle(H^j) = H^j$, that is,~$H^j$ is stable and has the structure described in Remark~\ref{remark:stable}. If~$H^{j+1} = \triangle(H^{j+1})$, the claim is true. Thus assume~$H^{j+1} \neq \triangle(H^{j+1})$. Remember that by the minimality of our counterexample, we have~$H^{i+1} = \triangle(H^i)$. By the definition of the triangle completion, the new edge~$e$ in~$H^{j+1}$ either connects two upper vertices or an upper and a lower vertex. Suppose first that~$e = \{v,v'\}$ with~$v,v' \in V_u$. Let~$U(v)$ and~$U(v')$ be the cliques containing~$v$ and~$v'$ among~$V_u$, respectively,  and let~$L(v)$ and~$L(v')$ be the vertices in~$V_\ell$ that are connected to~$U(v)$ and~$U(v')$, respectively. In~$\triangle(H^{j+1})$, the edges between~$v$ and~$U(v')$ and between~$v'$ and~$U(v)$ are inserted, as well as the edges between~$L(v)$ and~$v'$ and between~$L(v')$ and~$v$. In~$\triangle^2(H^{j+1})$, the graph induced by~$U(v) \dunion U(v')$ is rendered a clique and all edges between~$L(v)$ and~$U(v')$ and between~$L(v')$ and~$U(v)$ are inserted, resulting in a connected component that is complete bipartite between its upper and its lower vertices.

Suppose now that~$e = \{v,v'\}$ with~$v \in V_u$ and~$v' \in V_\ell$.
If~$v'$ is not adjacent to any upper vertex, then in~$\triangle(H^{j+1})$ all edges between~$v'$ and~$U(v)$ are inserted and~$\triangle(H^{j+1})$ is stable. Otherwise, let~$U(v')$ be the clique among~$V_u$ that is connected to~$v'$. Similarly as in the case that~$e$ is an edge among upper vertices, in~$\triangle^2(H^{j+1})$, the graph induced by~$U(v) \dunion U(v')$ is a clique and all edges between~$L(v)$ and~$U(v')$ are present. The vertices in~$L(v')$ are adjacent to all vertices in~$U(v')$ and to~$v$. A third triangle completion then accounts for the missing edges between~$L(v')$ and~$U(v)$.  

Therefore, the gap between two successive stable graphs is at most~$4$. 

For every~$i$ such that~$H^i$ is stable, the graph~$H^{i+1}$ contains an edge that connects two connected components from~$H^i$. We know that~$H^1$ is stable and we have shown that the gap between stable graphs is at most~$4$. Thus, also the gap between the graphs~$H^{i+1}$ that contain an extra edge connecting two connected components from~$H^i$ is at most~$4$. There are at most~$2m$ isolated vertices in~$H^1$, which yields~$k \leq 4 \cdot 2m = O(m)$.
\end{proof}

\begin{lem}\label{lem:triangle:in:next:it}
Let~$G$ be a graph that is completely cleaned up (i.\,e.,~$G$ satisfies~$\ccu(G) = G$). Then~$\Aux(G^{(1)}) \supseteq \triangle\big(\Aux(G)\big)$.
\end{lem}
\begin{proof}
Suppose~$G$ satisfies the assumption. Let~$V_u \dunion V_\ell$ be the partition of the vertices of~$\Aux(G)$ into upper and lower vertices. Let~$(C_1,M_1)$ and~$(C_2,M_2)$ be two vertices of~$V_u$ both adjacent to the vertex~$(D,N) \in V_\ell$. We prove that~$(C_1,M_1)$ and~$(C_2,M_2)$ are adjacent in~$\Aux(G^{(1)})$. 

By definition of the auxiliary graph, there exists a set of colors~$\mathcal{C}_1 \subseteq \mathcal{C}$ such that in~$G$, every vertex~$v\in C_1$ satisfies
\[
v \in M_1 \iff N^{+}_{\mathcal{C}_1}(v) = N.
\]

There is also a set~$\mathcal{C}_2\subseteq \mathcal{C}$  
such that in~$G$, every vertex~$v\in C_2$ satisfies
\[
v \in M_2 \iff N^{-}_{\mathcal{C}_2}(v) = N.
\]

Here, we use the fact that edge colorings respect converse equivalence.

For~$m\in M_1$ we have that~$m'\in M_2$ if and only if for every vertex~$w\in D$ it holds that~\[\chi(m,w)\in \mathcal{C}_1 \iff \chi(w,m')\in \mathcal{C}_2.\] Since~$G$ is completely cleaned up, this does not only hold for every~$w \in D$ but for every~$w \in V(G)$. This implies that in~$G^{(1)}$ there is a set of colors~$\mathcal{C'}$ such that every vertex~$v\in C_1$ satisfies
\[
v \in M_1 \iff N^{+}_{\mathcal{C}'}(v) = M_2.
\]

To see this, suppose~$v_1 \in M_1$. The new color of an edge~$(v_1, v_2)$ contains in its second component the multiset consisting of the tuples~$\big(\chi(w,v_2), \chi(v_1,w)\big)$ with~$w \in V(G)$. 

For~$v_2 \in M_2$ we have
\[\big(\chi(w,v_2), \chi(v_1,w)\big) \in \begin{cases}
\mathcal{C}_2 \times \mathcal{C}_1 &\text{if } w \in N \\
(\mathcal{C} \backslash \mathcal{C}_2) \times (\mathcal{C} \backslash  \mathcal{C}_1) &\text{if } w \notin N. \\
\end{cases}
\]

However, if~$v_1 \in M_1$ and~$v_2 \in (C_2 \backslash M_2)$, then the multiset in the new color of the edge~$(v_1, v_2)$ will contain some element from~$\big((\mathcal{C} \backslash \mathcal{C}_2) \times \mathcal{C}_1\big) \cup \big(\mathcal{C}_2 \times (\mathcal{C} \backslash \mathcal{C}_1)\big)$. 

Note that if~$v_2 \notin C_2$, then the new color of~$(v_1, v_2)$ is different from all colors of edges between~$C_1$ and~$C_2$ anyway since~$G$ is completely cleaned up. Thus, we obtain that in~$G^{(1)}$ the colors of edges from~$M_1$ to~$M_2$ are distinct from the colors of edges from~$M_1$ to the complement of~$M_2$.

By symmetry this shows that in~$\Aux(G^{(1)})$ the vertices~$(C_1,M_1)$ and~$(C_2,M_2)$ are adjacent. 

A similar argument shows that if~$(C_1,M_1)$ is adjacent to~$(C_2,M_2)$ and an upper or lower vertex~$(D,N)$, then in~$\Aux(G^{(1)})$ there is an edge between~$(C_2,M_2)$ and~$(D,N)$.

We conclude that~$\Aux(G^{(1)})\supseteq\triangle\big(\Aux(G)\big)$.
\end{proof}

After each complete clean-up step that Player~$2$ performs in the original game, she computes an auxiliary graph. We show that two auxiliary graphs differ by at least one edge if between their computations, in the original game at least one (vertex or edge) color class that is only incident with small vertex color classes is refined---either by a move of Player~$1$ or during the subsequent complete clean-up step performed by Player~$2$.

\begin{lem}\label{lemma:inequal_aux}
 Let~$G,G'$ with~$G \succeq G'$ be two graphs appearing in the game that have been completely cleaned up. If there is a color class in~$G$ that has been refined in~$G'$ and is only incident with small vertex color classes, then it holds that~$\Aux(G) \subsetneqq \Aux(G')$. 
\end{lem}

\begin{proof}
 Suppose after a complete clean-up step, a small vertex color class of~$G$ has been split into at least two new small vertex color classes~$C'$ and~$C''$ in~$G'$. Thus, because of Condition~\ref{item:c1} of a completely cleaned-up graph, the graph~$\Aux(G')$ contains an edge between the upper and the lower copy of~$(C',C')$, whereas none of these two vertices is present in~$\Aux(G)$.

 Now suppose that after a complete clean-up step, an edge color class~$C$ of~$G$ only incident with small vertex color classes has been split in~$G'$ into at least two new edge color classes~$C'$ and~$C''$ without causing any splitting of small vertex color classes. Consequently, there exists a vertex~$v$ in a small vertex color class~$A$ of~$G'$ which satisfies that~$\emptyset \subsetneqq N^+_{C'} (v) \subsetneqq N^+_C(v)$. Now we define~$M \coloneqq \{u \in A \mid N^+_{C'}(u) = N^+_{C'}(v)\}$. Let~$N$ be the vertex color class in~$G$ containing~$N^+_{C'}(v)$. The edge~$\big\{(A,M),\big(N,N^+_{C'}(v)\big)\big\}$ is present in~$\Aux(G')$ but not in~$\Aux(G)$.
\end{proof}

We wish to bound the costs that Player~2 incurs within Algorithm~\ref{algo}. To do so, we need to bound the costs that incur before and within the while loop. Let~$G_1,G_2,\ldots$ be the sequence containing the following two types of graphs in their order of appearance during the game. On the one hand, it contains the graphs that are played by Player~1 and cleaned up by Player~2 and have incurred some refinement of edges only incident with small vertex color classes. On the other hand, it contains the graphs that are results of computations of Line~\ref{line:computed:cleaned:graph} of Algorithm~\ref{algo} across the entire play of the game. Thus, the sequence consists of all completely cleaned-up graphs that are played in the game right after a refinement of color classes that are only incident with small vertex color classes.

\begin{lem}\label{lemma:sequence}
 The sequence of graphs~$G_1, G_2, \dots$  has length at most~$O(2^{t(n)}n)$.
\end{lem}

\begin{proof}
We set~$m \coloneqq 4n \cdot 2^{t(n)}$ for the threshold function~$t$. 
Each of~$\Aux(G_1), \Aux(G_2), \dots$ has size at most~$m$. Thus, since~$G_{i+1} \preceq G_{i}$, it holds that~$\Aux(G_{i})\subseteq \Aux(G_{i+1})$ by Remark~\ref{rem:aux}. 
 The while loop in Algorithm~\ref{algo} is only entered in the case that~$\triangle\big(\Aux(G)\big) \neq \Aux(G)$. Also in that situation the graph~$G$ is completely cleaned up. We conclude with Lemma~\ref{lem:triangle:in:next:it} that~$\triangle\big(\Aux(G_{i})\big) \subseteq \Aux(G_{i+1})$. With Lemma~\ref{lemma:inequal_aux} we obtain that~$\Aux(G_{i})\subsetneqq \Aux(G_{i+1})$.

Therefore, the sequence of the auxiliary graphs~$\Aux(G_1), \Aux(G_2), \dots$ fulfills the conditions of Lemma~\ref{lemma:aux} with~$m = O(2^{t(n)}n)$, implying that the length of the sequence~$G_1,G_2\ldots$ is~$O(m) = O(2^{t(n)}n)$.
\end{proof}

The lemma in particular implies that Algorithm~\ref{algo} terminates. Now we can bound the iteration number for small color classes.

\begin{cor}\label{cor:small:color:classes}
 The number of iterations in which color classes that are only incident with small vertex color classes are refined is~$O(2^{t(n)}n)$.
\end{cor}

\begin{proof}
 By Lemma~\ref{lemma:sequence}, the sequence of the computed auxiliary graphs satisfies the conditions from Lemma~\ref{lemma:aux}. Thus, every subsequence of it also satisfies these conditions.  If we only consider the subsequence of auxiliary graphs that are computed after moves that refine a color class that is only incident with small vertex color classes, we know that the sequence has length~$O(m)$, where~$m$ is the maximum number of upper vertices in an auxiliary graph. Thus, the sequence has length~$O(n \cdot 2^{t(n)})$.
 
 With Lemma~\ref{lemma:inequal_aux}, the length of the sequence is an upper bound on the number of iterations in which color classes that are only incident with small vertex color classes are refined. 
\end{proof}

From the corollary we immediately conclude Lemma~\ref{lem:linear:for:bounded:col:class}.

We have assembled all the required tools to prove our main theorem concerning the upper bound on the number of iterations.

\begin{proof}[Proof of Theorem~\ref{thm:bd:col:class:main:res}]
Let~$G$ be a graph with~$n$ vertices. To show a bound on the number of iterations, it suffices to show an upper bound on the costs of the 2-player game that we defined.
Assume that both players play optimally. By Corollary~\ref{cor:large:color:classes}, the total costs of moves in which a color class incident with a large vertex color class is split is~$O\big(n^2/t(n)\big)$. By Corollary~\ref{cor:small:color:classes}, the total costs of moves in which a small color class is split are~$O(n\cdot 2^{t(n)})$. Therefore, the entire game costs~$O\big(n^2/t(n)\big)+ O(n\cdot 2^{t(n)})$ and setting~$t \coloneqq \log_2(n)/2$ yields the upper bound of~$O\big(n^2/\log_2(n)\big)$.
\end{proof}

\section{Logics without counting}

The bound we have proven for the 2-dimensional Weisfeiler-Leman algorithm leads to bounds on the quantifier depth in the~$3$-variable fragment of first-order logic with counting~$\mathcal{C}^3$. However, one may wonder what happens for the 3-variable fragment of first-order logic without counting~$\mathcal{L}^3$. A priori it is not clear that there should be a relationship between the depths of formulas distinguishing graphs in~$\mathcal{C}^3$ and in~$\mathcal{L}^3$. 
We can thus not draw conclusions about~$\mathcal{L}^3$ from our theorems. 

However, a careful analysis of our proof reveals that we could have obtained the same results for the logic without counting.

For this, one has to redefine the coloring in Definition~\ref{def:k:dimensional:weisfeiler:lehman:refinement} so as to obtain a non-counting version of the 2-dimensional Weisfeiler-Leman algorithm (see \cite[Subsection 2.2.4]{Otto2017}). In this version, we replace the multiset~$\mathcal{M}$ with a set with the same elements. Refinement and stability is then defined with respect to this new operator. Most lemmas apply to the new situation with verbatim proofs. 
Most notably for the potential function for large color classes, the definition does not use a multiset. 

Condition~\ref{item:c2} in the clean-up steps of Player~2 then has to be replaced by a condition which requires that for~$v_1,v_2\in C_1$ we have that~$\{\chi (v_1,u)\mid u\in C_2\} = \{\chi (v_2,u)\mid u\in C_2\}$ and also~$\{\chi (u,v_1)\mid u\in C_2\} = \{\chi (u,v_2)\mid u\in C_2\}$. 

Finally, we highlight the fact that in the auxiliary graph, the definition of adjacency does not require counting either. For example, the existence of a set of colors~$\mathcal{C}' \subseteq \mathcal{C}$ such that every vertex~$v$ satisfies
 \[
 v \in M \iff N^{+}_{\mathcal{C}'}(v) = N,
 \]
can obviously be expressed in a~$\mathcal{C}^2$-formula without counting quantifiers. We conclude that analogous statements to our theorems also hold for the 3-variable first-order logic without counting~$\mathcal{L}^3$.

\section{Conclusion}

We have shown that the number of iterations of the 2-dimensional Weisfeiler-Leman algorithm is in~$O\big(n^2/\log(n)\big)$. 

The factor~$1/\log(n)$ arises from a trade-off between considerations in large and small vertex color classes, and thus it arises from the factor of~$2^t$ appearing in Lemma~\ref{lem:linear:for:bounded:col:class}.
However, an improvement of the bound on the iteration number for graphs with bounded color class size would not directly improve the overall iteration number. Indeed, since refinements within small color classes may be caused by refinements of large color classes, and also since small color classes might appear only over time, we crucially needed to bound the number of all refinements involving small color classes within general graphs (that may not have bounded color class size). This is exactly what our game is suitable for, since Player~1 may continue to refine the graph after it is stable. 
 
Our proof for the 3-variable logic already requires a careful analysis of the interaction between the small color classes. However, it remains an interesting open question whether our techniques can be generalized to also show bounds on the depth of formulas with more variables.

\appendix
\section{Converse equivalence}\label{app:converse:equiv}

In this section we briefly discuss the quite technical reason why we require converse equivalence for all colorings.
Indeed we required that~$\chi(v_1,u_1) = \chi(v_2,u_2)$ if and only if~$\chi(u_1,v_1) = \chi(u_2,v_2)$ for any coloring~$\chi$ that we consider (see the preliminaries).
In principle it would be possible to avoid such a definition. However, consider the following example.
Let~$A \coloneqq \{a_1,a_2,\ldots,a_t\}$ and~$B \coloneqq \{b_1,b_2,\ldots,b_t\}$ be two sets of equal size. We define a coloring~$\chi$ such that for all~$a,a'\in A$ and~$b,b'\in B$ with~$a \neq a'$ and~$b \neq b'$ we have that~$\chi(a,a) = \chi(b,b) = 0$,~$\chi(a,a') = 1$,~$\chi(b,b')= 2$,~$\chi(b,a) = 3$.
We also define~$\chi(a_i,b_j) =4$ if~$i = j$ or~$i = j+1$ modulo~$t$ and define~$\chi(a_i,b_j) =5$ otherwise. Note that this coloring does not have the converse equivalence property.
If we apply the definition of the Weisfeiler-Leman algorithm given in Definition~\ref{def:k:dimensional:weisfeiler:lehman:refinement}, then this coloring is stable. For such a coloring one has to adapt the multiset~$\mathcal{M}$ to also take reverse directions into account and replace the definition by
\[{\mathcal{M}} \coloneqq \big\{\!\!\big\{\big(\chi(w,v_2),\chi(v_2,w), \chi(v_1,w) ,\chi(w,v_1)
\big) \mid w\in V     \big\}\!\!\big\}.\]
In the logic context, especially when working with finite structure this is the natural definition for the Weisfeiler-Leman algorithm (see for example~\cite{KieferSS15}). Also in the context of coherent configurations, this is always required~\cite{MR1994960}. Since the definition ensures that an iteration of the Weisfeiler-Leman algorithm can also take reverse directions into account, after one iteration the coloring satisfies converse equivalence. (Recall that loops have different colors than other edges.) Converse equivalence is then maintained in all future iterations as well. 

In the game we consider in this paper, we basically want to allow Player 1 to refine to arbitrary colorings that do not arise from the application of the algorithm. Indeed, we can drop the requirement for converse equivalence. In this case, at the expense of a more technical clean-up procedure, Player 2 could then (under the new definition of the Weisfeiler-Leman algorithm) restore converse equivalence. Overall, we obtain essentially the same results.

\bibliographystyle{alpha}
\bibliography{paper_arxiv_update}

\end{document}